\documentclass[11pt,a4paper]{article}
\usepackage{amsfonts,amsmath,amsthm,epsfig,verbatim,amssymb,amscd,eucal,bbm}

\numberwithin{equation}{section}
\newtheorem{prop}{Proposition}[section]

\newtheorem{cor}[prop]{Corollary}
\newtheorem{lem}[prop]{Lemma}
\newtheorem{thm}[prop]{Theorem}

 \newcommand{\mrm}[1]{\text{\rm #1}}

 \newcommand{\A}{\mathcal{A}}
 \newcommand{\D}{\mathcal{D}}
 \newcommand{\cR}{\mathcal{R}}
 
 \newcommand{\fE}{\mathfrak{E}}

 \newcommand{\R}{\mathbb{R}}
 \newcommand{\1}{\mathbbm{1}}
 \newcommand{\C}{\mathbb{C}}
 \newcommand{\E}{\mathbb{E}}
\newcommand{\p}{\mathbb{P}}
  \newcommand{\Z}{\mathbb{Z}}
 \newcommand{\N}{\mathbb{N}}

 \newcommand{\Lom}{\mathcal{L}}

 \def\eref#1{(\ref{#1})}
\begin{document}

\noindent

\title{\bf Services within a Busy Period of an M/M/1 Queue and Dyck
  Paths}

\author{Moez {\sc Draief}
\thanks{LIAFA, Universit{\'e} Paris 7, case 7014,
2 place Jussieu, 75251 Paris Cedex 05, France.
E-mail: {\tt Moez.Draief@liafa.jussieu.fr}
and {\tt  Jean.Mairesse@liafa.jussieu.fr}.}
\and
Jean {\sc Mairesse}\,\footnotemark[1]
}

\maketitle

\begin{abstract}

We analyze the service times of customers in a stable M/M/1 queue in
equilibrium depending on their position in a busy period.
We give the law of the service of a customer at the beginning, at the end,
or in the middle of the busy period. It enables as a by-product to prove
that the process of instants of beginning of services is not Poisson.
We then proceed to a more precise analysis. We consider a family of
polynomial generating series associated with Dyck paths of length $2n$
and we show that they provide the correlation function of the
successive services in a busy period with $n+1$ customers.

\end{abstract}

\renewcommand\abstractname{R\'esum\'e}
\begin{abstract}
On s'int\'eresse \`a l'analyse des temps de service des clients
d'une file M/M/1 stable et en \'equilibre selon
leur position dans une p\'eriode d'activit\'e.
On donne la loi d'un service sachant que le client se
trouve au d\'ebut, \`a la fin ou au milieu de la p\'eriode d'activit\'e. Ceci
permet, au passage, de prouver que le processus des instants de
d\'ebut de service n'est pas un processus de
Poisson. On m\`ene ensuite une \'etude plus fine.
On exhibe une famille de s\'eries g\'en\'eratrices
polyn\^omiales associ\'ees aux chemins de Dyck de longueur $2n$  et on
montre qu'il s'agit de la fonction de corr\'elation des diff\'erents services dans une
p\'eriode d'activit\'e comportant $n+1$ clients.
\end{abstract}

\smallskip


\smallskip

{\noindent\bf AMS classification (2000):} 60K25, 68R05.


\section{Introduction}

The M/M/1/$\infty$/FIFO queue (or M/M/1 queue) is the queue with a
Poissonian arrival stream, exponential services, a single server,
an unlimited buffer capacity, and a First-In-First-Out service
discipline. It can be argued that the M/M/1 queue is
the most elementary and the most studied system in queueing theory,
see for instance \cite{cohe82,taka,robe,ScWe}. Quoting \cite{ScWe},
``most likely, any book with {\em queueing} in the title has something
to say on the subject''.

Let $\lambda$ be the intensity of the Poisson arrival process and let
$\mu$ be the parameter of the exponential service times. Assume that
the stability condition $\lambda < \mu$ holds and consider the queue
in equilibrium. Our objective is to get precise information on the
distribution of the service of a customer based on its position  in
the busy period.

First of all, recall that the distribution of the first, respectively
last,  service is an
exponential of parameter $\mu$, respectively $(\lambda+\mu)$.
We are then able
to compute the distribution of a service in the ``middle''
of a busy period (i.e. neither at the beginning nor at the end).
As a by-product, we also get the distribution of the duration between
two successive beginning of services. Since it is not an exponential,
we conclude that the point process of
the instants of beginning of services is not Poisson (as opposed to the point process
of completion of services).

Then we study the service time of the $k$-th customer in a busy period of length $n+1$
(i.e. containing $n+1$ customers). Consider a busy period conditionned to be of length $n+1$,
and let $(q_i)_{0,\dots,2n+2}$ be the corresponding embedded queue-length excursion.
Its trajectories are equiprobable and it is easy to see that they are in bijection with Dyck paths of length $2n$.
If we condition $(q_i)_{0,\dots,2n+2}$ to be associated with a given Dyck path $\pi$ of length $2n$ then we observe
that the law of the service time of the $k$-th customer is equal to the convolution product of $l_k+1$ exponentials
of parameter $\lambda+\mu$ where $l_k$ is the length of the intersection of $\pi$ with the line $y=x-2k$.
By summing over Dyck paths of length $2n$, we get an expression for the joint law of the services in a busy period of length $n+1$.
Then using elementary properties of Dyck paths, we obtain results on services within a busy period somewhat difficult to obtain
by direct probabilistic arguments (Section 4). The correlation function of the services is a natural generating polynomial of
Dyck paths following a simple integral recursion (Section 5).

Using the combinatorial properties of lattice paths to study the busy period of simple queues is classical,
see \cite{Flajolet,Guillemin, taka2} and references therein.
In these articles, quantities such that the area swept by the queue-length process during a busy period are studied, with a much more involved
combinatorial analysis than what is presented below for the sequence of services within a busy period. This should come as no surprise.
The area and related quantities, are derived by counting in a Dyck path the number of ascents and descents of a given vertical coordinate
(Dyck paths are lattice paths in $\N\times \N$, see Section 3). On the contrary, the sequence of services is derived by counting in a Dyck path
the number of ascents of a given horizontal coordinate (roughly speaking). This is in essence like working with generating polynomials of Dyck paths in
{\em non-commuting} variables. It is therefore hopeless to get as precise information.

\section{In the Middle of the Busy Period}
Given a positive real random variable $X$ with law $\mu$, denote its Laplace transform by
$\Psi_{X}(s) = \Psi_{\mu}(s)= \int \exp (-sx) d\mu(x), s\in \C, \mrm{Re}(s)\geq 0$.
We write $\Lom[X\mid \A]=\p\{X \in \cdot \:\mid \A\}$  for the conditional
law of $X$ given an event $\A$. The corresponding Laplace transform
is denoted $\Psi_{X\mid\A}(s)$. The convolution product of two probability distributions $\mu$ and $\nu$ is denoted by
$\mu \star \nu$. The
indicator function of a subset $A$ of a set is denoted by $\1_{A}$.
It is convenient to denote by $\mrm{Exp}(a)$ the exponential
distribution of parameter $a$ defined by $\mrm{Exp}(a) [x,+\infty) =
\exp (-ax), x\in \R_+$.
Recall that $\Psi_{\mrm{Exp}(a)}(s)= a/(a+s)$.

\medskip

We consider an $M/M/1$ queue with the following notations.
Let $(A_n)_{n\in \Z}$ be the arrival Poisson process of intensity
$\lambda$. Let $(\tau_n)_{n\in \Z}$ be the inter-arrival times, with
$\tau_n =A_{n+1}-A_n$. Denote by $(\sigma_n)_{n\in \Z}$ the service
times of the customers. The sequence $(\sigma_n)_{n}$ is
i.i.d. and $\sigma_0 \sim \mrm{Exp}(\mu)$.
We assume that the stability condition $\lambda < \mu$ is satisfied,
and we consider the queue in equilibrium. Let $(Q_t)_{t\in \R}$ be the
queue-length process, where $Q_t$ is the number of customers either in
service or in the buffer at time $t$.

The state of the server can be described as an alternating sequence of
idle and busy periods. A {\em busy period}
is a maximal period during which $Q_t>0$. An idle period is a maximal
period during which $Q_t=0$.
The {\em length} $|B|$ of a busy period $B$ (not to be confused with its
duration) is the number of customers served during the busy period.
Throughout, when we consider a generic busy period $B$, we denote for
simplicity by $\sigma_0, \dots, \sigma_{|B|-1}$ and $\tau_0, \dots,
\tau_{|B|-1}$ respectively the service times and the inter-arrival
times of the different customers in the busy period.

\begin{lem}\label{le-easy}
Let $\Delta_n$ be  the event that a generic busy period consists
of $n+1$ customers, then
\begin{equation}
\Delta_n =\{\sum_{j=0}^i \tau_j < \sum_{j=0}^i \sigma_j,\:
i=\{0,\dots,n-1\};\:\sum_{j=0}^n \tau_j  \geq \sum_{j=0}^n
\sigma_j\}\:.
\end{equation}
\end{lem}
\noindent
The justification is easy.

The durations of successive busy periods and idle
periods are independent random variables.
The duration of an idle period is clearly distributed as
$\mrm{Exp}(\lambda)$. The distribution of a busy period is more
complex. The next results can be found for instance in \cite[Chapter
II.2.2]{cohe82} or \cite[Chapter 1.2]{taka}.
The probability that a busy period $B$
consists of $(n+1)$ customers is given by
\begin{equation}\label{eq-length}
\p\{|B|=n+1\} = C_{n}
\frac{\lambda^{n}\mu^{n+1}}{(\lambda+\mu)^{2n+1}}\:,
\end{equation}
where $C_n$ is the $n$-th Catalan number, see \S \ref{se-dyck}.
Let $\delta_n$ be the conditional law of the duration of a busy period, given that the length
of the busy
period is $(n+1)$.
The Laplace
transform of $\delta_n$ is given by
\begin{equation}\label{duration-B}
\Psi_{\delta_n}(s) =\frac{(\lambda+\mu)^{2n+1}}{(\lambda+\mu+s)^{2n+1}}\:.
\end{equation}
Hence, $\delta_n$ is the distribution of the sum of $(2n+1)$ i.i.d. r.v.'s of law
$\mrm{Exp}(\lambda+\mu)$.

\medskip

Given two independent random variables $X \sim \mrm{Exp}(\alpha)$
and $Y \sim \mrm{Exp}(\beta)$, where $\alpha,\:\beta \in \R^*_+$,
recall that
\begin{equation}\label{eq-recall}
\Lom[X\mid X\geq Y] = \mrm{Exp}(\alpha +\beta) \star  \mrm{Exp}(\alpha) , \:
\Lom[X\mid X < Y] = \mrm{Exp}(\alpha +\beta)\:.
\end{equation}
Using elementary arguments based on the memoryless property of the
exponential distribution, we get:
\begin{equation}\label{eq-basic}
\sigma_0 \sim \mrm{Exp}(\mu), \ \sigma_{|B|-1} \sim \mrm{Exp}(\lambda+\mu)
\:.
\end{equation}
Furthermore, remarking that $\{|B|=1\}=\{\sigma_0 \leq \tau_0\}$ and
using \eref{eq-recall}, it follows that:
\begin{equation}\label{eq-basic2}
\Lom[\sigma_0 \mid |B|=1] = \mrm{Exp}(\lambda+\mu),\:
\Lom[\sigma_0 \mid |B|>1 ] = \mrm{Exp}(\lambda+\mu)
\star \mrm{Exp}(\mu)\:.
\end{equation}
Our goal is now to derive the law of a service in the {\em middle} of
$B$, i.e. of a service which is neither the first nor the last one
(assuming that $|B|>2$).

Let $\sigma_*$ be the service of a generic customer numbered $*$
and let $B$ be the busy period it belongs to. Define the events
\begin{eqnarray*}
\fE_{\mrm{o}} &=& \{* \mrm{ is the {\bf o}nly customer of } B\} \ = \ \{ |B|=1\} \\
\fE_{\mrm{f}} &=& \{* \mrm{ is the {\bf f}irst customer of } B \mrm{ and } |B|>1\} \\
\fE_{\mrm{l}} &=& \{* \mrm{ is the {\bf l}ast customer of } B \mrm{ and } |B|>1\} \\
\fE_{\mrm{m}} &=& \{* \mrm{ is in the {\bf m}iddle of } B \mrm{ and } |B|>2\}\:.
\end{eqnarray*}
Clearly the four events are disjoint and $\p\{\fE_{\mrm{o}} \cup \fE_{\mrm{f}} \cup
\fE_{\mrm{l}} \cup \fE_{\mrm{m}}\}=1$. Since the lengths of successive busy periods
are i.i.d., we obtain immediately that
\begin{eqnarray*}
\p\{\fE_{\mrm{f}} \} &=&  \p\{\fE_{\mrm{l}} \} \\
\p\{\fE_{\mrm{f}} \} &=& \p\{|B|>1\} /\E[|B|] \\
\p\{\fE_{\mrm{o}} \cup \fE_{\mrm{f}}\} &=& 1/\E[|B|] \:.
\end{eqnarray*}
Now using (\ref{eq-length}), we get $\p\{|B|>1\}=\lambda/(\lambda+\mu)$
and $\E[|B|]=\mu/(\mu-\lambda)$. It follows that
\begin{equation}\label{eq-proba}
\p\{\fE_{\mrm{o}}\}= \frac{\mu-\lambda}{\mu+\lambda}, \ \ P\{\fE_{\mrm{f}}\}=
\p\{\fE_{\mrm{l}}\}=\frac{\lambda(\mu-\lambda)}{\mu(\mu+\lambda)},\: \p\{\fE_{\mrm{m}}\}=
\frac{2\lambda^2}{\mu(\mu+\lambda)}\:.
\end{equation}
Clearly, $\sigma_0 \sim \Lom[\sigma_*\mid \fE_{\mrm{o}} \cup
\fE_{\mrm{f}}]$ and $\sigma_{|B|-1} \sim \Lom[\sigma_*\mid
\fE_{\mrm{l}}]$. We deduce that
\begin{eqnarray*}
\Psi_{\sigma_*}(s) & = & \p\{\fE_{\mrm{o}} \cup \fE_{\mrm{f}}\}
\Psi_{\sigma_*\mid \fE_{\mrm{o}} \cup
\fE_{\mrm{f}}}(s) +\p\{\fE_{\mrm{l}}\} \Psi_{\sigma_* \mid
\fE_{\mrm{l}}}(s) + \p\{\fE_{\mrm{m}}\} \Psi_{\sigma_*\mid \fE_{\mrm{m}}}(s) \\
& = & \p\{\fE_{\mrm{o}} \cup \fE_{\mrm{f}}\} \Psi_{\sigma_0}(s) +
\p\{\fE_{\mrm{l}}\} \Psi_{\sigma_{|B|-1}}(s) + \p\{\fE_{\mrm{m}}\} \Psi_{\sigma_*\mid \fE_{\mrm{m}}}(s) \:.
\end{eqnarray*}
That is,
\begin{equation*}
\frac{\mu}{\mu + s} =  \frac{\mu-\lambda}{\mu} \frac{\mu}{\mu+s} +
\frac{\lambda(\mu-\lambda)}{\mu(\mu+\lambda)}
\frac{\mu+\lambda}{\mu+\lambda+s} +
\frac{2\lambda^2}{\mu(\mu+\lambda)} \Psi_{\sigma_*\mid \fE_{\mrm{m}}}(s) \:.
\end{equation*}
After simplification of the above expression, we obtain the
Laplace transform of the conditional law of $\sigma_*$ on the event
$\fE_{\mrm{m}}$ :
\begin{equation}\label{eq-midservice}
\Psi_{\sigma_*\mid \fE_{\mrm{m}}}(s)= \frac{(2\mu + s)(\mu + \lambda)}{2(\mu +s)(\mu + \lambda
+s)}\:.
\end{equation}

As a by-product, we can prove that the process of
instants of beginning of services is not a Poisson process,
in contrast
with the process of completion of services (departure instants) which
is Poisson of intensity $\lambda$ according to Burke Theorem
\cite{burk,reic}.
Let us detail the argument. Let $\xi_*$ be the difference between the
instants of beginning of services of two generic successive customers
numbered $*$ and $(*+1)$. Let $B$ be the busy period of $*$ and let $\nu$ be the first idle period following
$B$ ($\nu\: \sim \: \mrm{Exp}(\lambda))$. Using \eref{eq-basic},
\eref{eq-basic2}, we get immediately that
\begin{eqnarray*}
\xi_* \1_{\fE_{\mrm{o}}}=(\sigma_0+\nu) {\1}_{\fE_{\mrm{o}}} &
\Longrightarrow \ \
\Lom[\xi_*\mid \fE_{\mrm{o}}] & = \ \ \mrm{Exp}(\lambda+\mu) \star
\mrm{Exp}(\lambda)\\
\xi_* \1_{\fE_{\mrm{f}}}=\sigma_0 {\1}_{\fE_{\mrm{f}}} &
\Longrightarrow \ \
\Lom[\xi_*\mid \fE_{\mrm{f}}]& = \ \ \mrm{Exp}(\lambda+\mu) \star
\mrm{Exp}(\mu)\\
\xi_* {\1}_{\fE_{\mrm{l}}}=(\sigma_{|B|-1}+\nu) {\1}_{\fE_{\mrm{l}}} &
\Longrightarrow \ \
\Lom[\xi_*\mid \fE_{\mrm{l}}]& = \ \ \mrm{Exp}(\lambda+\mu) \star
\mrm{Exp}(\lambda)\:,
\end{eqnarray*}
and
\begin{equation*}
\Lom[\xi_* \mid \fE_{\mrm{m}}]  =  \Lom[\sigma_* \mid \fE_{\mrm{m}}]\:.
\end{equation*}
Since we have just computed $\Psi_{\sigma_* \mid \fE_{\mrm{m}}}(s)$, we
deduce the Laplace transform of $\xi_*$:
\begin{equation}\label{eq-interbeg}
\Psi_{\xi_*}(s) = \frac{\lambda (\mu^2(\lambda+\mu) +
  \mu(2\mu+\lambda)s+ \lambda
  s^2)}{\mu(\lambda+s)(\mu+s)(\lambda+\mu+s)}\:.
\end{equation}
We check on this expression that $\E[\xi_*]=1/\lambda$ and we have
$\E[\xi_*^2]
=\frac{\lambda^3-\mu\lambda^2+\mu^3+\mu^2\lambda}{\lambda^2\mu^2(\lambda
  +\mu)}$. In particular, we have $\E[\xi_*^2] < \E[d_*^2] =
2/\lambda^2$, where $d_*$ is a generic inter-departure time.

\section{Dyck Paths}
\label{se-dyck}

The Catalan numbers $(C_n)_{n\in \N}$ are defined by
\begin{equation}\label{eq-catalan}
C_n = \frac{1}{2n+1} {2n+1 \choose n}=\frac{1}{n+1} {2n\choose n}\:.
\end{equation}
The generating function of these numbers is given by
\begin{equation*}
\sum_{n=0}^{+\infty}C_nx^n=\frac{1-\sqrt{1-4x}}{2x}\:.
\end{equation*}
The first Catalan numbers are
$C_0=1,\:C_1=1,\:C_2=2,\:C_3=5,\:C_4=14,\:C_5=42
,\:C_6=132,\:C_7=429,\cdots$. They appear in many combinatorial
contexts see for instance \cite{GKPa,stan2}. In particular, $C_n$
is the number of Dyck paths of length $2n$. A {\em Dyck path}  of
length $2n$ is a  path in the lattice $\N \times \N$ which begins
at the origin $(0,0)$ ends at $(0,2n)$ and with steps of type
$(1,1)$ or $(1,-1)$. Denote by $\D_n$ the set of Dyck paths of
length $2n$, observe that $\D_0$ is a singleton whose element is
the unique Dyck path of length $0$.

We now define a family of polynomials
related to Dyck paths.
Let $\pi \in \D_n$ and let
$\gamma_j$ be the line $y=x-2j$, for $j\in \{0,\dots,n-1\}$ and denote by
$\alpha_j$ the length of the intersection of $\gamma_j$ with $\pi$
(equivalently $\alpha_j+1$ is the number of lattice points common
to $\pi$ and $\gamma_j$). We introduce two polynomials $P_{\pi}$
and $R_{\pi}$ defined by
\begin{equation}
P_{\pi}(y_0,y_1,\dots,y_{n-1}) = \prod_{i=0}^{n-1}
\frac{y_i^{\alpha_i}}{\alpha_i !}\:,\:\:\:\:\:
R_{\pi}(y_0,y_1,\dots,y_{n-1}) = \prod_{i=0}^{n-1}
y_i^{\alpha_i}\:.
\end{equation}
Let $(P_n)_{n \in \N}$  and $(R_n)_{n \in \N}$ be the two families
of polynomials defined by $P_n =\sum_{\pi \in \D_n} P_{\pi},\:R_n =\sum_{\pi\in \D_n} R_{\pi}$.
Clearly $P_n$ and $R_n$ are homogeneous polynomials of degree $n$
over the $n$ variables $y_0,y_1,\dots,y_{n-1}$.

\section{The Law of the Services in a Busy Period}
Recall that the queue-length process $(Q_t)_{t\in\R}$ is a continuous time Birth-and-Death process on $\N$ with
generator $P$ such that $P_{n,n+1}=\lambda,\:n\geq 0\:;P_{n,n-1}=\mu,\:n\geq
1\:;P_{n,m}=0,\:|n-m|\geq 2$. Let $(q_n)_{n\in\Z}$ denote the Markov chain embedded at its jump instants.
More precisely, let $T$ be the point process obtained as the superposition  of the arrival and departure processes and let $(T_n)_{n\in\Z}$
 be its points with the convention $T_0=A_0$. Then we set $q_n=Q_{T_n^-}$. The transition matrix of $(q_n)_{n\in\Z}$ is given by
\begin{equation}\label{transition}
p_{0,1}=1;\:p_{i,i-1}=\frac{\mu}{\lambda+\mu};\:p_{i,i+1}=\frac{\lambda}{\lambda+\mu},\:i\geq1;
\end{equation}
and $p_{i,j}=0$ otherwise.

\begin{figure}[h]
\[\epsfxsize=280pt \epsfbox{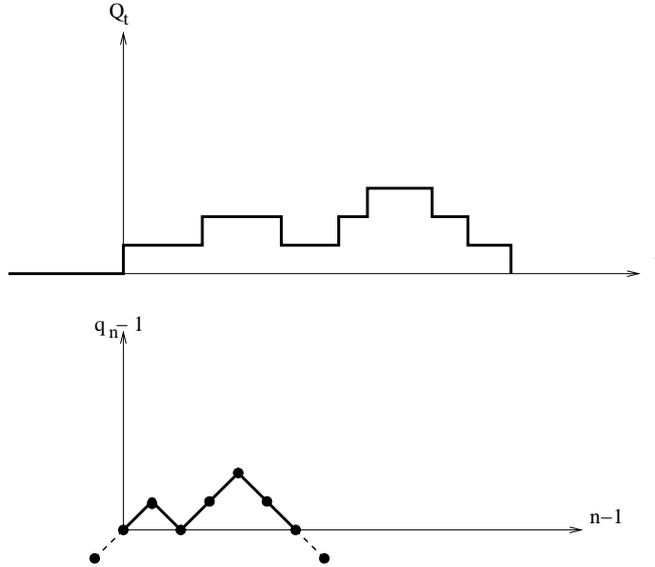} \]
\caption{Dyck path associated with a busy period.}
\label{bijec}
\end{figure}

A busy period corresponds to an excursion of $(q_n)_{n\in\Z}$ from $0$ to its first return to $0$.
With the same numbering convention as in Section $2$,
the generic busy period $B$ consists of $n+1$ customers if and only if
\begin{equation}
q_0=0,\:q_i>0,\: i \in \{1,\cdots,2n+1\},\:q_{2n+2}=0\:.
\end{equation}
On this event, the (random) path with  successive edges
$(i-1,q_i-1),\:i\in\{1,\cdots,2n+1\}$ is a (random) Dyck path of
length $2n$. We call it the Dyck path {\em associated with} $B$
(see  Figure \ref{bijec}). On the event $\{|B|=n+1\}$, all Dyck
paths appear with the same probability (the probability of a given
trajectory $(q_n)_n$ depends only on the number of increasing and
decreasing jumps, see (\ref{transition})). On the event that
$|B|=n+1$ and that the associated Dyck path is $\pi \in \D_n$, the
power of $y_{i-1}$ in $P_{\pi}$ is the number of customers which
join the system between the the $i$-th and the $(i+1)$-th
departures. Combining these observations with the fact that the
time between successive transitions of $(Q_t)_{t\in\R}$ are
independent r.v.'s of law $\mrm{Exp}(\lambda+\mu)$ as long as the
queue is non empty, we get :

\begin{thm}\label{main}
Given that the length of the busy period is $n+1$,
the conditional density of the random vector
$(\sigma_0,\dots,\sigma_{n})$ representing the service
times of the successive customers is
\begin{equation}
D(y_0,\dots,y_n)=\frac{(\lambda+\mu)^{2n+1}}{C_n} e^{- (\lambda + \mu) (y_0+\dots+y_{n-1})}
P_n(y_0,\dots,y_{n-1})e^{-(\lambda + \mu) y_n}\:,
\end{equation}
where $P_n$ is the Dyck polynomial of
degree $n$ defined in Section 3.
\end{thm}
A direct computation of the Laplace transform leads to the following :
\begin{cor}\label{Laplace trans}
Consider a random vector
$(\tilde{\sigma_0},\dots,\tilde{\sigma_{n}})\sim\Lom[(\sigma_0,\dots,\sigma_n)\mid|B|=n+1]$.
Its Laplace transform is given by
\begin{eqnarray}
\Psi_{(\tilde{\sigma}_0,\dots,\tilde{\sigma}_{n})}(s_0,\dots,s_n)&=& \E [ \prod_{i=0}^n
e^{-s_i \tilde{\sigma}_i} ]\\\label{Laplace}
&=&\frac{1}{C_n}(\prod_{i=0}^{n-1} z_i) R_n(z_0,\dots,z_{n-1})z_n\:,
\end{eqnarray}
where $z_i = \frac{\lambda+\mu}{ \lambda + \mu + s_i},\: \forall i \in
\{0,\dots,n\}$, and $R_n$ is defined in Section 3.
\end{cor}
Let us paraphrase the above results in a somewhat more intuitive way.
In a busy period of length $(n+1)$, the conditional law of
$(\sigma_0,\dots \sigma_{n})$ is the same as the law of
$(\tilde{\sigma}_0,\dots \tilde{\sigma}_{n})$ that we now describe.
The law of $\tilde{\sigma}_{n}$ is an $\mrm{Exp}(\lambda+\mu)$ independent of
$(\tilde{\sigma}_0,\dots ,\tilde{\sigma}_{n-1})$. Let $\Pi$ be a
r.v. uniformly distributed over $\D_n=\{\pi_1,\dots
,\pi_{C_n}\}$. Conditionally on $\{\Pi=\pi_i\}$, the r.v.'s $\tilde{\sigma}_j$
are independent and distributed as the sum of $k_j^i$ random variables of law
$\mrm{Exp}(\lambda+\mu)$, where $(k_j^i-1)$ is the exponent of $y_j$
in $P_{\pi_i}$. This is illustrated in Table 1.

We now exploit the correspondance with Dyck paths.

Let $\D_n^{<i>}$ be the set of Dyck paths of length $2n$ where the
first return to the axis $\{(n,0),\: n \in \N\}$, after the origin
$(0,0)$, occurs at the point $(2i,0)$, $i\in \{1,\dots,n\}$. Clearly, the sets  $\D_n^{<i>}$
are disjoint and $\D_n= \cup_{i=1}^{n}\D_n^{<i>}$. Furthermore
\begin{equation}\label{bijection}
\D_n^{<i>} \simeq \D_{i-1} \times \D_{n-i}\:.
\end{equation}

A consequence of the above is the very classical identity on Catalan
numbers :
\begin{equation*}
C_n=\sum_{i+j=n-1}C_iC_j\:.
\end{equation*}
Let $\cR_n$ be defined by $\cR_n(z_0,\dots,z_{n-1})=R_n(z_0,\dots,z_{n-1})\times z_0 \dots z_{n-1}$, going back to Corollary \ref{Laplace trans}, we have
\begin{equation*}
\Psi_{(\tilde{\sigma}_0,\dots,\tilde{\sigma
}_{n})}(s_0,\dots,s_n)=\frac{1}{C_n}\cR_n(z_0,\dots,z_{n-1}).z_n\:.
\end{equation*}
where $z_i = \frac{\lambda+\mu}{ \lambda + \mu + s_i},\: \forall i \in
\{0,\dots,n\}$. We also define $\cR_n^{(i)}(z_i)=\cR_n(1,\dots,1,z_{i},1,\dots,1)$, then
\begin{equation*}
\Psi_{\tilde{\sigma}_i} (s_i) =\frac{1}{C_n}\cR_n^{(i)}(z_i)\:.
\end{equation*}
\begin{prop}\label{Rec}
On the event $\{|B|=n+1\}$, we have
\begin{equation}
\cR_0^{(0)}(z_0)=z_0\:,\forall n \geq 0\:,\:\cR_n^{(0)}(z_0)=\sum_{i+j=n-1}z_0\cR_i^{(0)}(z_0)C_j\:,
\end{equation}
for, $0< k \leq n-1$,
\begin{equation}
\cR_k^{(k)}(z_k)=C_k z_k\:,\:\cR_n^{(k)}(z_k)=\sum_{i+j=n-k-1}\cR_i^{(k)}(z_k)C_j +
\sum_{l=0}^{k-1}\cR_{n+l-k}^{(l)}(z_k) \:.
\end{equation}
\end{prop}
\begin{center}
\begin{figure}[hbt]
\[\epsfxsize=250pt \epsfbox{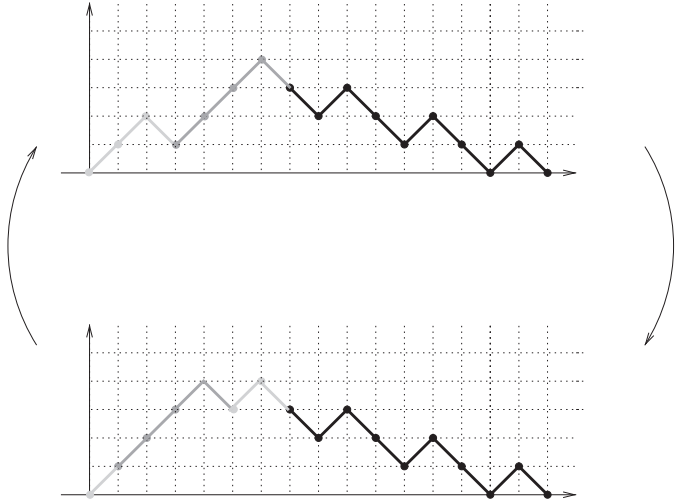} \]
\caption{The mapping $\Psi\: :\: \D_n \longrightarrow \D_n.$}
\label{Psi}
\end{figure}
\end{center}
On Table 1, one notices a simple relation between the laws of
$\sigma_0$ and $\sigma_1$, which is actually always true :
\begin{prop}
Let $B$ be a generic busy period, for $n \geq 1$ we have
\begin{equation}
\Lom[\sigma_0 \mid |B|=n+1]=\Lom[\sigma_1 \mid |B|=n+1]\star \mrm{Exp}(\lambda+\mu)\:.
\end{equation}
\end{prop}
\begin{center}
\begin{tabular}{|c|c|c|c|c|c|c|c|c|}
\hline
&&&&&&&& \\
&$\mrm{Exp}(\lambda+\mu) $& 1 & 2 & 3 & 4 & 5 & 6 & 7  \\
&&&&&&&& \\ \hline
 $|B|=4$ &$\sigma_0$ & & 2 & 2 & 1 & & & \\
 &$\sigma_1$ & 2 & 2 & 1 & & &&  \\
 &$\sigma_2$ & 3 & 2 & & & &&  \\
  &$\sigma_3$ & 1 & & & & & & \\ \hline
 $|B|=5$ &$\sigma_0$ & & 5 & 5 & 3 & 1 &&  \\
 &$\sigma_1$ & 5 & 5 & 3 & 1 & &&  \\
 &$\sigma_2$ &7 &5 &2 & & && \\
 &$\sigma_3$ &9 &5 & & & & & \\
& $\sigma_4$ & 1 & & & & && \\ \hline
  $|B|=6$ &$\sigma_0$ & &14  &14  &9  &4  &1& \\
 &$\sigma_1$ &14  &14  &9  &4  &1 &&  \\ \hline
  $|B|=7$ &$\sigma_0$ & & 42 & 42 & 28 & 14 &5 &1  \\
 &$\sigma_1$ & 42 & 42 & 28& 14 &5 &1 &  \\  \hline
  \end{tabular}
$$\mbox{Table 1. Services.}$$
The table should be read as follows. For instance, on $\{|B|=5\}$, the law of $\sigma_2$ is $\mu_{\sigma_2}=\frac{7}{14} \mrm{Exp}(\lambda+\mu)+\frac{5}{14}\mrm{Exp}(\lambda+\mu)\star\mrm{Exp}(\lambda+\mu)
+\frac{2}{14}\mrm{Exp}(\lambda+\mu)\star\mrm{Exp}(\lambda+\mu)\star\mrm{Exp}(\lambda+\mu)$.
\end{center}

\begin{proof}
The mapping $\Psi : \D_n \longrightarrow \D_n$ is defined in Figure
\ref{Psi}. It is clearly an involution, hence a bijection.
More formally, given a Dyck path $\pi \in \D_n$ such that
$R_{\pi}=y_1^k.y_0Q(y_0,y_2,\dots,y_{n-1})$ then $\Psi(\pi)\in \D_n$ is
defined by $R_{\Psi(\pi)}=y_0^{k+1}Q(y_1,y_2,\dots,y_{n-1})$. In view
of Corollary \ref{Laplace trans}, it completes the proof.
\end{proof}

\section{Dyck Paths Polynomials}
We go back to the family of polynomials $(P_n)_{n \in \N}$ defined in Section
\ref{se-dyck}. We are going to use Theorem \ref{main} to give nice expressions for the $P_n$'s.
Let $\Delta_{n}$ be the event that a generic busy period consists of
$n+1$ customers.
Let $A_0,\dots,A_n$ be borelians of $\R_+^*$,
\begin{equation*}
\p\{\sigma_i \in A_i,\: i=0,\dots ,n \mid \Delta_n\} = \frac{\p\{\sigma_i
\in A_i,\: i =0,\dots ,n ; \Delta_n\}}{\p\{\Delta_n\}}\:.
\end{equation*}
Let $L_n =\p\{\sigma_i \in A_i,\: i \in\{0,n\};\Delta_n\}$ and for
$k=0,\dots,n$, let $Y_k = \sum_{i=0} ^{k}y_i$ and $X_k = \sum_{i=0}
^{k}x_i$. Using Lemma \ref{le-easy}, we have
\begin{eqnarray*}
L_n&=&(\lambda \mu)^{n+1}\int_{A_0 \times \dots \times A_n}e^{-\mu
Y_n}\: dy_0 \dots dy_n \int_0^{Y_0}e^{-\lambda x_0} dx_0\int_0^{Y_1-X_0}e^{-\lambda
x_1}dx_1\cdots \\
&&\hspace*{6cm} \cdots \int_0^{Y_{n-1}-X_{n-2}}e^{-\lambda x_{n-1}}
dx_{n-1}\int_{Y_n-X_{n-1}}^{\infty}e^{-\lambda x_n} dx_n\\
&=&\lambda^{n} \mu^{n+1} \int_{A_0\times \dots\times A_n}e^{- (\mu +
  \lambda)Y_n}\:dy_0\dots dy_n
\int_0^{Y_0} dx_0 \int_0^{Y_1-X_0}
dx_1\cdots\int_0^{Y_{n-1}-X_{n-2}}dx_{n-1}\:.
\end{eqnarray*}
Then, using theorem \ref{main}, we get
\begin{equation}\label{iterint1}
P_n(y_0,\dots,y_{n-1})=\int_{0}^{y_0}dx_0\int_{0}^{y_0+y_1-x_0}dx_1\dots\int_{0}^{y_0+\dots+y_{n-1}-(x_0+\dots+x_{n-2})}dx_{n-1}\:.
\end{equation}
Simple manipulations of formula (\ref{iterint1}) then yield :
\begin{lem}\label{recursion}
The polynomials $(P_n)_{n \in \N}$ satisfy the following equations
\begin{equation}\label{recpoly}
P_n(y_0,\dots,y_{n-1})=\int_{y_1}^{y_0+y_1}
P_{n-1}(y,y_2,\dots,y_{n-1})\:dy
\end{equation}
and
\begin{equation}\label{recpoly2}
P_n(y_0,\dots,y_{n-1}) =\int_{0}^{y_0}\:dx_0
 \int_{0}^{y_1+x_0}\:dx_1\:\dots\int_{0}^{y_{n-1}+x_{n-2}}\:dx_{n-1}\:.
\end{equation}
\end{lem}
For completeness, here is a direct proof of (\ref{recpoly}) without using Theorem \ref{main}.

Let $\D_n^{(i)}$ be the set of all Dyck paths of length $2n$ starting
with $i$ steps of type $(1,1)$ followed by one step of type $(1,-1)$
and define the polynomial $P_n^{(i)}$ such that
\begin{equation*}
\sum_{\pi \in \D_n^{(i)}}P_{\pi}=\frac{y_0^i}{i!}P_{n}^{(i)}(y_1,\dots,y_{n-1})\:.
\end{equation*}
Clearly, we have
\begin{equation*}
P_n(y_0,\dots,y_{n-1})= \sum_{i=1}^{n}\frac{y_0^i}{i!}P_{n}^{(i)}(y_1,\dots,y_{n-1})\:.
\end{equation*}
\begin{figure}[h]
\[\epsfxsize=480pt \epsfbox{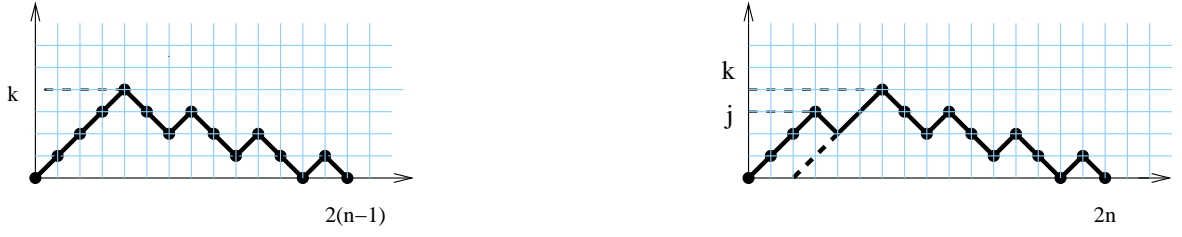} \]
\caption{Proof of the equality (\ref{extension}):  Paths contributing to $P_k^{(n-1)}$ (Left) and to $P_j^{(n)}$ (Right).}
\label{extend}
\end{figure}
Hence, we get
\begin{eqnarray*}
\int_{y_1}^{y_0+y_1} P_{n-1}(y,y_2,\dots,y_{n-1})\:dy &=& \int_{y_1}^{y_0+y_1}\sum_{i=1}^{n-1}\frac{y^i}{i!}P_{n-1}^{(i)}(y_2,\dots,y_{n-1})\:dy\\
&=&\sum_{i=1}^{n-1}\frac{1}{(i+1)!}
[(y_0+y_1)^{i+1}-y_1^{i+1}]P_{n-1}^{(i)}(y_2,\dots,y_{n-1})\\
&=&\sum_{j=1}^n
\frac{y_0^j}{j!}\sum_{k=0}^{n-j}\frac{y_1^k}{k!}P_{n-1}^{(k+j-1)}(y_2,\dots,y_{n-1})\\
\end{eqnarray*}
With the help of Figure \ref{extend}, we notice that
\begin{equation}\label{extension}
P_{n}^{(j)}(y_1,\dots,y_{n-1})=\sum_{k=j-1}^{n-1}\frac{y_1^{k-j+1}}{(k-j+1)!}  P_{n-1}^{(k)}(y_2,\dots,y_{n-1})=\sum_{k=0}^{n-j}\frac{y_1^k}{k!}  P_{n-1}^{(k+j-1)}(y_2,\dots,y_{n-1})\:.
\end{equation}
\begin{figure}[hbt]
\[\epsfxsize=350pt \epsfbox{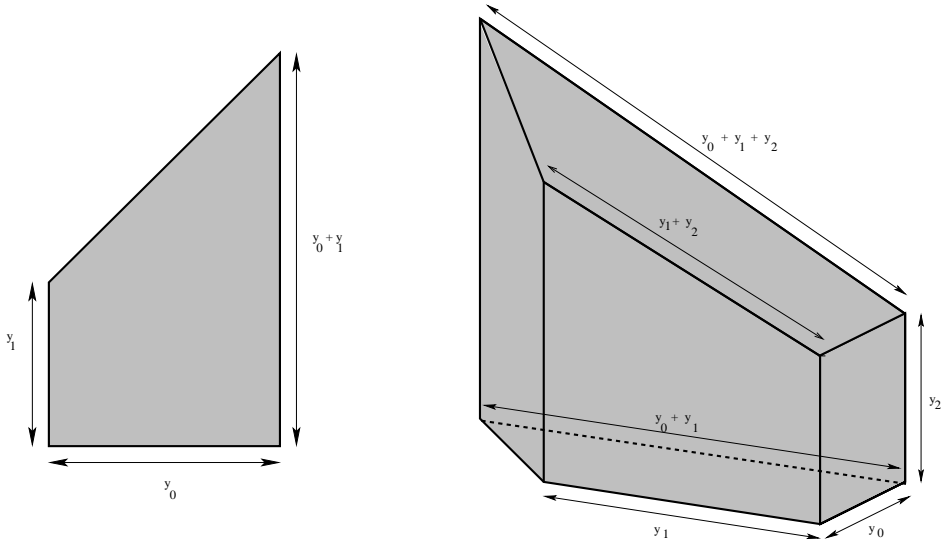} \]
\caption{The volumes of the gray areas are $P_2(y_0,y_1)$ (left) and
  $P_3(y_0,y_1,y_2)$ (right).}
\label{Volume}
\end{figure}

It leads to
\begin{equation*}
\int_{y_1}^{y_0+y_1} P_{n-1}(y,y_2,\dots,y_{n-1})\:dy=\sum_{i=1}^n \frac{y_0^i}{i!}P_n^{(i)}(y_1,\dots,y_{n-1})
= P_n(y_0,\dots,y_{n-1})\:.
\end{equation*}

This result can also be proved using the theory of species presented in \cite{Leroux}.
Finally, using (\ref{recpoly2}), the polynomials $(P_n)_{n\in \N}$ can be interpreted as volumes. We give a
representation of this in Figure \ref{Volume} for $n=2$ and $n=3$.

\paragraph{\bf Conclusion.}
Here are several other simple models of queues for which the queue-length process is a Birth-and-Death process: the M/M/K/$\infty$ queue,
the M/M/$\infty$ queue, or the M/M/K/L queue ($K \leq L < \infty$). In each case, if the generic busy period is of length $n+1$,
we can associate with it a Dyck path of length $2n$. However, the different Dyck paths of length $2n$ are not equiprobable anymore.
Hence, we do not get a simple formula for the joint law of the services as in Theorem \ref{main}.


\begin{thebibliography}{1}

\bibitem{Leroux}
F.~Bergeron, G.~Labelle and P.~Leroux,
\newblock Combinatorial species and tree-like structures.
\newblock Cambridge University Press, 1998.

\bibitem{burk}
P.~Burke,
\newblock The output of a queueing system.
\newblock Operations Research 4 (1956) 699-704.

\bibitem{cohe82}
J.W.~Cohen,
\newblock The single server queue. 2nd edition.
\newblock North-Holland, Amsterdam, 1982.


\bibitem{Flajolet}
P.~Flajolet and F.~Guillemin,
\newblock The formal theory of Birth-and-Death processes, lattice path combinatorics, and continued fractions.
\newblock Advances in Applied Probability 32 (2000) 750-778.

\bibitem{GKPa}
R.~Graham, D.~Knuth, and O.~Patashnik,
\newblock Concrete mathematics: a foundation for computer science. 2nd edition.
\newblock Addison-Wesley, 1994.

\bibitem{Guillemin}
F.~Guillemin and D.~Pinchon,
\newblock On the area swept under the occupation process of an M/M/1 queue in a busy period. Queueing Systems Theory
Appl. 29 (1998), no. 2-4, 383--398.

\bibitem{reic}
E.~Reich,
\newblock Waiting times when queues are in tandem.
\newblock Ann. Math. Stat. 28 (1957) 527-530.

\bibitem{robe}
P.~Robert,
\newblock R\'eseaux et files d'attente: m\'ethodes probabilistes.
\newblock Number~35 in Math\'ematiques \& Applications. Springer, 2000.

\bibitem{ScWe}
A.~Schwartz and A.~Weiss,
\newblock Large deviations for performance analysis. Queues,
  communications, and computing.
\newblock Chapman \& Hall, London, 1995.


\bibitem{stan2}
R.~Stanley,
\newblock Enumerative Combinatorics, Vol. 2.
\newblock Number~62 in Cambridge Studies in Advanced Mathematics. Cambridge
  University Press, 1999.

\bibitem{taka}
L.Tak\'{a}cs,
\newblock Introduction to the theory of queues.
\newblock University Texts in the Mathematical Sciences. Oxford University
  Press, 1962.

\bibitem{taka2}
L.~Tak\'{a}cs,
\newblock Queueing methods in the theory of random graphs,
\newblock Probability and Stochastics Series, CRC, Boca Raton, FL, (1995) 45-78.

\end{thebibliography}
\end{document}